\documentclass[letterpaper]{aipproc}
\layoutstyle{6s}

\def\be{\begin{equation}}
\def\ee{\end{equation}}
\def\bea{\begin{eqnarray}}
\def\eea{\end{eqnarray}}
\def\ba{\begin{array}}
\def\ea{\end{array}}

\def\part{\partial}

\def\cH{\cal H}

\def\hi{\hat \imath}

\begin{document}

\title[Quantum Myers effect for D0/D4 systems]{Quantum Myers effect and its Supergravity dual for D0/D4
systems}
\author{Pedro J. Silva}{
address={Physics Department, Syracuse University, Syracuse, New York 13244},
email={psilva@physics.syr.edu}
}

\begin{abstract}
The presence of a distant D4-brane is used to further investigate
the duality between M-theory and D0-brane quantum mechanics.
A polarization of the quantum mechanical ground state is found.
A similar deformation arises for the bubble of normal space found near D0-branes
in classical supergravity solutions.
These deformations are compared and are shown to have the same structure in
each case.

\end{abstract}

\date{DTP-yy-nn}

\maketitle

In recent years the outlook within string theory has changed immensely.
The new cornerstones of the theory are non-perturbative duality conjectures.
Some of the most impressive such conjectures are those of
matrix theory \cite{bfss} and Maldacena's AdS/CFT conjecture
\cite{malda}.

These conjectures relate the physics of certain gravitating systems to
that of specific non-gravitating gauge theories.
The dynamics of the dual field theories are deduced
from the low energy effective actions of the various non-abelian D-brane
systems. The correspondences appear to rely on the
particular form of the non-abelian interactions.

Recently, several investigations \cite{tvr2,mye1}
have uncovered the form of certain non-abelian couplings of D-branes
to supergravity background fields. Our goal here is to invest the
`polarization' of the Dp-brane bound state in the background
of a D(p+4)-brane.  For definiteness, we shall concentrate on the D0/D4
context.

In certain cases, the application of a Ramond-Ramond background field
to a D0-brane system induces a classical dielectric effect
and causes the D0-branes to deform into a non-commutative D2-brane
\cite{mye1}.  While the Ramond-Ramond fields of our D4-brane background will
not be strong enough to induce such a classical effect, they do modify
the potential that shapes the non-abelian character of the
quantum D0 bound state.  As a result, this bound
state is deformed, or polarized. Fundamental to this studies will be
the connection described by Polchinski
\cite{pol1} relating the size of the matrix
theory bound state to the size of the bubble of space that is well-described
by classical supergravity in the near D0-brane spacetime.

The near D0-brane spacetime is obtained by taking a limit in which
open strings decouple from closed strings and the result is a
ten-dimensional spacetime which has
small curvature and small string coupling when one is reasonably
close (though not too close) to the D0-branes.  However, if one moves
beyond some critical $r_c$, the curvature reaches the string scale.
As a result, the system beyond $r_c$ is not adequately described by
the massless fields of classical supergravity.

Our goal is therefore to compare the deformations of the non-abelian D0-brane
bound state with the deformations of this bubble of `normal' space around
a large stack of D0-branes. As has become common
in string theory, we find that the quantum mechanical effects of the non-abelian
D0-brane couplings correctly reproduce the effects of classical supergravity in
the large $N$ limit.

Throughout this document a series of approximation and large N calculations
together with hand-waving arguments will
be given, with no justification, the actual justifications can be found on
\cite{ms}.

We begin with the world-volume
effective field theory describing $N$ D0-Branes in the standard
D4-brane background. This action is a suitable
generalization of the action for a single D0-brane, consisting of
the Born-Infeld term together with appropriate Chern-Simon terms.
However, the full action encodes the Chan-Paton factors or non-abelian
degrees of freedom that arise from strings stretching between the D0-branes.

The first part of the non-abelian D0 effective action
is the Born-Infeld term
\bea
S_{BI}=-T_0 \int dt \, STr\left( e^{-\phi}
\sqrt{-\left( P\left[E_{ab}+E_{ai}(Q^{-1}-\delta)^{ij}E_{jb}
\right]\right) \, det(Q^i{}_j)} \right)
\label{eq:1}
\eea
with $E_{AB} = G_{AB}+ B_{AB}$ and $Q^i{}_j\equiv\delta^i{}_j + i\lambda\,[\Phi^i,\Phi^k]\,E_{kj}\ .
\label{eq:2}$
In writing (\ref{eq:1})
we have used a number of conventions taken from Myers \cite{mye1}:

The rest of the action is given by the
non-abelian Chern-Simon terms
\bea S_{CS}=\mu_0\int dt
STr\left(P\left[e^{i\lambda\,\hi_\Phi \hi_\Phi} ( \sum
C^{(n)}\,e^B)\right]\right)\ .
\eea
The symbol $i_\Phi$ is a non-abelian generalization of the
interior product with the coordinates $\Phi^i$, $i_\Phi
\left(\frac{1}{2}C_{AB}dX^AdX^A\right) = \Phi^iC_{iB}dX^B.$

The D4-brane background is defined by the metric $G_{AB}$, the dilaton
$\phi$, and the Ramond-Ramond 6-form field strength $F_{A_1A_2A_3A_4A_5A_6}$:
\bea
\label{D4}
&&ds^2_4 = {\cal H}_4^{-1/2} \eta_{\mu\nu}dX^\mu dX^\nu +
{\cal H}_4^{1/2}\delta_{mn}dX^{m}dX^{n} \nonumber \\
&&e^{-2\phi} = {\cal H}_4^{1/2} \;\;\;,\;\;F_{01234m} = \partial_m {\cal H}_4^{-1},
\eea
with all other independent components of the field strength vanishing.
Here the space-time coordinates described by the index
$A$ have been partitioned into directions parallel to the D4-brane
(which we will label with a Greek index $\mu$) and directions
perpendicular to the D4-brane (which we label with a Latin index
$m$).  The function ${\cal H}_4 = 1 + (r_4/|X|)^3$
is the usual harmonic function of the D4-brane solution with
 $|X|^{\,2}=\delta_{mn}X^mX^n$ and with $r_4 = (gN_4)^{1/3}  l_s$
being the constant that sets the length scale of the supergravity
solution.

Expanding the Born-Infeld and the Chern-Simon action in this background
we get,
\bea
\label{SwF} S_{eff.} = -T_0\lambda^2 \int dt \, STr\left\{
{1\over {2 g_{tt}}}\partial_t\Phi\partial_t \Phi +
{1\over 4}[\Phi,\Phi]^2 + \right.\hspace{4cm}&& \nonumber \\
\left. + \lambda \left( {1\over 2}\partial_t\Phi^i\partial_t\Phi^j\Phi^k\partial_k(g_{tt})^{-1}+
{1\over {2 g_{tt}}}\partial_t\Phi^i\partial_t\Phi^j\Phi^k\partial_kg_{ij}\;+ \right.\right. \hspace{1cm}&&\nonumber \\
\left. \left.+{1\over 2}[\Phi,\Phi^i][\Phi^j,\Phi]\Phi^k\partial_k g_{ij}+ \frac{1}{10}\Phi^{i}\Phi^{j}\Phi^{k}\Phi^{l}\Phi^{m}F_{\tau
ijklm}\right) \right\}.&&
\eea
The are also Fermion terms that are rather long, and little insight is gained
by writing them explicitly here.

One might begin with a discussion of classical solutions corresponding to
the above effective action.  However, aside from the trivial commutative
solution, one does not expect to find any static solutions.However, aside from the trivial commutative solution, one
does not expect to find any static solutions\footnote{The
literature \cite{indios,gjs,clt} contains some examples of
classical non-commutative solutions in similar (but non-supersymmetric)
systems.}. Nevertheless, we may expect that the non-abelian couplings to the
background affect the quantum bound state by altering the shape of the
potential and thus the ground state wavefunction. Let us calculate the size of the
ground state by considering the expectation value of the squared radius
operator $R^2 \equiv \lambda^2Tr(\Phi^2) = \lambda^2 Tr(\Phi^i \Phi^j g_{ij})$.
Note that by passing to an orthonormal frame one finds a full SO(9) spherical
symmetry in the $O(\lambda^2)$ terms in our action.  Thus, to $O(\lambda^2)$
the expectation value of $Tr (\Phi^i \Phi^i g_{ii})$ is independent of $i$
and $R^2$ is the radius of the corresponding sphere measured in terms
of string metric proper distance.

Here we give a simple argument for the behavior of $\langle R^2 \rangle$
based on the usual 't Hooft scaling behavior.
Our strategy is to treat the couplings to the D4-brane fields as perturbations
to the D0-brane action in flat empty spacetime.
Thus, we divide (\ref{SwF}) into an `unperturbed action' $S_0$ and a
perturbation $S_1$. Note that as we place the N D0-branes far
from the D4-branes, the Ramond-Ramond coupling term can be written in the form
\begin{equation}
\Phi^{\mu_1}\Phi^{\mu_2}\Phi^{\mu_3}\Phi^{\mu_4}\Phi^m
F_{0 \mu_1 \mu_2 \mu_3 \mu_4 m} =
\frac{f}{\lambda^{1/2}}\Phi^{\mu_1}\Phi^{\mu_2}\Phi^{\mu_3}\Phi^{\mu_4}
\epsilon_{\mu_1 \mu_2 \mu_3 \mu_4} \Phi^{m} \frac{X^m}{|X|}
\end{equation}
where $f
= 3 (r_4^3\lambda^{1/2})/(z_\perp^4)
\approx 3{\cal H}_4^{-2}(r_4^3\lambda^{1/2}/(z_\perp^4)$
is a scalar dimensionless measure of the field strength.
Here $z_\perp$ is the distance between the N D0-branes and the D4-brane, and
$\epsilon_{\mu_1 \mu_2 \mu_3 \mu_4}$ is the antisymmetric symbol on four
indices.  In what follows we
treat all effects of the D4-brane only to lowest order in $({\cal H}_4-1)$
and $f = 3 (r_4^3\lambda^{1/2})/(z_\perp^4)$, so that $f
\approx {\cal H}_4^{-2}f$. With this understanding, the other $O(\lambda^3)$
terms are also proportional to $f$.

It will be useful to express the dynamics in terms of
rescaled fields and a rescaled time coordinate:
\be
\tilde \Phi^i = \lambda^{1/2} {\cal H}_4^{1/12}
 (gN)^{-1/3} \Phi^i\;\;,\;\;
\tilde t = \lambda^{-1/2}
{\cal H}_4^{-1/3} (gN)^{1/3}t. \ee This yields the action
\cite{ch}
\begin{equation}
\label{FermAct}
S_0 = - N \int d\tilde t \,STr \left( - \frac{1}{2} \partial_{\tilde t} \tilde
\Phi \partial_{\tilde t} \tilde \Phi + \frac{1}{4} [\tilde \Phi, \tilde \Phi]^2
\right)
\end{equation}
and the perturbation
\bea
&&S_1=- \left[ (gN)^{1/3}  {\cal H}_4^{-1/12} f \right]
N\int d\tilde t\,STr\left({1\over 10}\tilde\Phi^{i}\tilde\Phi^{j}\tilde\Phi^{k}\tilde\Phi^{l}\tilde\Phi^{m} \epsilon_{
ijklm}+ \right. \nonumber \\
&&\hspace{0cm}\left. {1\over 2}\partial_t\tilde\Phi^i\partial_t\tilde\Phi^j\tilde\Phi^k
\frac{ \partial_k(g_{tt})^{-1}}{f}+{1\over {2}}\partial_t\tilde\Phi^i\partial_t\tilde\Phi^j\tilde\Phi^k
\frac{ \partial_kg_{ij}}{f}+
{1\over 2}[\tilde\Phi,\tilde\Phi^i][\tilde\Phi^j,\tilde\Phi]\tilde\Phi^k
\frac{\partial_k g_{ij}}{f} \right),\nonumber \\
&&\equiv - [(gN)^{1/3} {\cal H}_4^{-1/12}  f]\,\tilde S_1 .
\eea
Note that in writing
$\tilde S_1$ we have extracted a factor of $f$ from $S_1$.  The advantage
of this form is that both $S_0$ and $\tilde S_1$ are
manifestly independent of $g$, $\lambda$, and
$f$ while they depend on $N$ only through the overall factor and the trace.
The dependence of $S_0$ and $\tilde S_1$
on ${\cal H}_4$ is only though contractions with $g_{ij}$.
These could be further eliminated by passing to an orthonormal frame, so
the dynamics of
scalar contractions such as $\Phi^i\Phi^j g_{ij}$ will be independent of
${\cal H}_4$.

Let us now consider the case $f=0$ and the corresponding ground state
$\langle R^2 \rangle_0$.
We will think of this as the limit of small $\frac{\ell_s}{z_\perp}$, so
that we preserve ${\cal H}_4 \neq 1$.
Note that we have
\begin{equation}
\label{unpert}
\langle R^2 \rangle_0 = (gN)^{2/3} {\cal H}_4^{-1/6} \lambda
\langle Tr \tilde \Phi^2 \rangle_0.
\end{equation}
The factor $\langle Tr \tilde \Phi^2 \rangle_0$ is manifestly independent
of $g$, ${\cal H}_4$, and $\lambda$,
and the form of $S_0$ is the usual one associated with 't Hooft scaling
for which $\langle Tr \tilde \Phi^2 \rangle_0$ is also independent of $N$
in the limit of large $N$ with $gN$ fixed. This reproduces the results
of \cite{pol1,bfss}:
\begin{equation}
\label{qmr0}
\sqrt{\langle R^2 \rangle_0} \sim (gN)^{1/3} {\cal H}_4^{-1/12} \lambda^{1/2},
\end{equation}
where the product of $(g {\cal H}_4^{-1/4})^{1/3}$ can be viewed as the
natural dependence on the local string coupling $ g_{local} \equiv g e^\phi
= g{\cal H}^{-1/4}_4$ of the D4-brane background.

Let us now turn to the perturbed system. Considering the ground
state expectation value as the low temperature
limit of a thermal expectation value gives a Euclidean path integral
for $\langle R^2 \rangle$.  We wish to expand the factor $e^{-S_1} =
e^{-\left((gN)^{1/3} {\cal H}_4^{-1/12}  f \right) \tilde S_1}$ as
$1 - (gN )^{1/3} {\cal H}_4^{-1/12}  f  \tilde S_1 + (gN)^{2/3}
{\cal H}_4^{-1/6} f^2 \tilde S_1^2
- ...$.  Note that the order zero term gives just
$\langle R^2 \rangle_0$, the expectation value in the unperturbed
ground state.  The contribution from the
first order term then vanishes because
$\tilde S_1$ is anti-symmetric under a total inversion of space while
$R^2$, $S_0$, and the integration measure are invariant.
Thus, the leading contribution is of second order
in $\tilde S_1$, in the 't Hooft limit we express our final result as
\begin{equation}
\label{qmres}
\frac{\langle R^2 \rangle -  \langle R^2 \rangle_0}{\langle R^2 \rangle_0}
\sim (gN)^{2/3}  {\cal H}_4^{-1/6} f^2.
\end{equation}
where we have isolated the dependance on $g,N,f$.

Having considered the quantum mechanical description of the non-abelian
D0-brane bound state,
we now wish to compute a corresponding effect in classical
supergravity. We seek a BPS solution containing both D0's and D4's
with the D0's being both fully localized and separated from the
D4-branes.  It is conceptually simplest to discuss the full D0/D4
solution and then take a suitable decoupling limit.
Such full solutions are known exactly, but only as an
infinite sum over Fourier modes \cite{mar1}. As a result, we find
it more profitable here to follow a perturbative method as suggested by the
quantum mechanical calculation above.  We therefore expand
the supergravity
solution in $f$, the magnitude of the Ramond-Ramond
4-form field strength at the location of the zero-branes.

Let us consider a BPS system of D4-branes and $N$
D0-branes with asymptotically flat boundary conditions. Using
the usual isotropic ansatz in the appropriate
gauge reduces the problem to solving the equations
\cite{HM,Tseytlin,BREJS,AIRV}
\bea
\left(\partial_{\bot}^2+{\cal H}_4\partial_{\|}^2\right)
      {\cal H}_0 = 0\;\;\;,\;\;\;
{\cal H}_4=1+\left(\frac{r_4}{|X|}\right)^3, \label{eq:6}
\eea
where as before the D4-brane lies at $X^m=0$,
$|X|^{\,2}=\delta_{mn}X^mX^n$, and
${\cal H}_4$ and ${\cal H}_0$ are the `harmonic' functions for the
D4-brane and D0-brane respectively.
The two relevant derivative operators are a flat-space Laplacian
($\partial_{\|}^2 \equiv \sum^{\mu=4}_{\mu=1}\partial_\mu\partial_\mu$)
associated with the directions parallel to the D4-brane and
another ($\partial_{\perp}^2 \equiv
\sum^{m=9}_{m=5}\partial_m\partial_m$) associated with the perpendicular
directions.

In order to treat the D4-branes as a perturbation, we place them
far away from the D0-branes. It is convenient to change to new
coordinates $x^m$ (lowercase) whose origin is located at the D0
singularity.  One of these coordinates is distinguished by running
along the line connecting the D0- and D4-branes.  Let us call this
coordinate $x_\perp$.  The other four $x^m$ coordinates will play
a much lesser role. Introducing the distance $z_{\perp}$ between
the D0- and  D4-branes and expanding ${\cal H}_4$ to first order
about the new origin yields \bea {\cal H}_4 \approx {\cal
H}_4(x=0) - 3\left( \frac{r_4}{z_\perp} \right)^3
\left(\frac{x_{\perp}}{z_\perp}\right) \equiv {\cal H}_4(0) +
\delta{\cal H}_4. \label{eq:cs} \eea Here we have used $z_\perp\gg
(r_4,x_\perp)$, since  the D4-branes are located far away. Note
that fixing $z_\perp$ sets the location of the D0 singularity
relative to the D4-brane. However, as we will see, it is
not clear in general that the position of the singularity
corresponds precisely to the center of mass. Equation \ref{eq:6}
can be solved by expanding ${\cal H}_0$ in terms of
$\delta{\cH}_4\,$ i.e. $\,{\cH}_0={\cH}_{00}+{\cH}_{01}+... $
where ${\cal H}_{0n}=O(\delta {\cal H}_4^{\,\,\,n})$. We find
\bea
\label{eqtos}
(\partial^2_\perp + {\cal H}_4(0)
\partial_\parallel^2) {\cH}_{00}= 0, \;\; &{\rm so \ that}&
\;\;{\cH}_{00}=1+\left(\frac{r_0}{r}\right)^7, \nonumber \\
(\partial^2_\perp + {\cal H}_4(0) \partial_\parallel^2) {\cH}_{01}=
\delta{\cH}_4\partial^2_\parallel{\cH}_{00} \;\;&{\rm and}&\;\;
(\partial^2_\perp + {\cal H}_4(0) \partial_\parallel^2) {\cH}_{02}=
\delta{\cH}_4\partial^2_\parallel{\cH}_{01}
\eea
Here we have introduced $r^2= |x|^2 \equiv \delta_{mn}x^m
x^n+ {\cal H}_4^{-1}(0)
\delta_{\mu \nu}x^\mu x^\nu$, a sort of coordinate distance from the D0-brane.
Note that this $r$ does in fact label spheres of symmetry for the
unperturbed solution ${\cal H}_{00}$.

Since the D4-brane background has altered the background metric for the
D0-brane system, this will change certain familiar normalizations.
We therefore note that total electric flux from the D0-branes
must equal the number $N$ of D0-brane charge quanta
$(g \ell_s^7)/(60\pi^3).$
Since the D4-brane is far away, we may compute this flux in a
region where ${\cal H}_0 \approx 1$ but where ${\cal H}_4 = {\cal H}_4(0)$.
The result yields $(gN\ell_s^7)/(60 \pi^3) = {\cal H}_4^2 r_0^7.$

The above equations (\ref{eqtos})
are easily solved in terms of Green's functions. We stress that
under a rescaling of coordinates
$y,x \rightarrow \beta y, \beta x$ the function ${\cal H}_{01}$ scales
homogeneously as $\beta^{-6}$.  As a result, we may write
\be
{\cH}_{01} = \frac{\omega_1 f}{x^6} r_0^7 \lambda^{-1/2},
\ee
where $\omega_1$ is an unknown dimensionless function of the
angles associated with the direction cosines $x^A/r$ and
$f=(3\lambda^{1/2}r_4^3)/(z_\perp^4)$. Furthermore,
${\cal H}_{01}$ is even under any $x^\mu \rightarrow x^\mu$ and under
any $x^m \rightarrow x^m$ for $x^m \neq x_\perp$.  However, we see that
${\cal H}_{01}$ is odd under $x_\perp \rightarrow - x_\perp$. Thus,
the ${\cal H}_{01}$ term provides an (angle dependent) {\it shift}
of the bubble so that it is no longer centered on the D0-brane center of mass.
In particular, this has no effect on the {\it size} of the bubble.
For the second order term, we have
\be
{\cH}_{02} = \frac{\omega_2 f^2}{x^5},
\ee
where $\omega_2$ is again a dimensionless function of the angles.
This time, however, ${\cal H}_{02}$ is even under $x^A \rightarrow x^A$
for any $A$.  As a result, ${\cal H}_{02}$ directly encodes a change in
the size of the bubble.

Let us now calculate the size of this solution. We follow Polchinski
\cite{pol1} and use the measure that successfully reproduces the size
of the unperturbed D0-brane bound state. This means that we should locate
the surface enclosing the D0-brane singularity where the string metric
is so strongly curved that it has structure on the string scale. Inside
this surface is a bubble of space that is well described by classical
supergravity.  However, when $r$ is large
the proper distance $2\pi {\cal H}_4^{1/4} {\cal H}_0^{1/4} r$
around the sphere enclosing the origin may still be on the order of
$\ell_s$ so that the metric clearly has structure on the string scale.
In particular, one might think of strings encircling
the bubble itself.  The region inside this surface is to be
associated with the quantum D0-bound state, and one expects the bubble of
`normal' space and the bound state to have corresponding sizes and shapes.

Now, given the correspondence between the $r$ of supergravity
isotropic coordinates and the non-abelian $\Phi^2$ in D0 quantum mechanics
in the absence of the D4-brane, it is natural to expect a correspondence
between the
$R^2 = Tr (\Phi^i \Phi^j g_{ij})$ on the QM calculation and
the supergravity quantity $R^2 = {\cal H}_4^{1/2} \delta_{mn}x^m x^n
+ {\cal H}_4^{-1/2} \delta_{\mu \nu}x^\mu x^\nu = {\cal H}_4^{1/2} r^2$.
The edge of the bubble lies at the value of $R \equiv {\cal H}_4^{1/4} r$
for which $\ell_s \sim r {\cal H}_0^{1/4} {\cal H}_4^{1/4}
 = R {\cal H}_0^{1/4}(r).$  This yields the relation
\bea
\label{Reqn}
R \sim \ell_s [{\cH}_0(r)]^{-1/4} \sim
\ell_s {\cH}_{00}^{-1/4}\left(
1-\frac{{\cH}_{01}}{4{\cH}_{00}}-\frac{{\cH}_{02}}{4{\cH}_{00}}+
\frac{5}{20}\left(\frac{{\cH}_{01}}{4{\cH}_{00}}\right)^2 +
O(f^3)\;\right).
\eea

In order to make connections with the quantum mechanical
calculations, we wish to consider this system in the decoupling
limit $g \rightarrow 0$ with fixed $gN$ and $r/\ell_s \sim g^{1/3}
\rightarrow 0$.  Note that this scaling may seem more familiar when expressed in
terms of the eleven-dimensional plank mass $M_{11} = g^{-1/3} \ell_s^{-1}$
as it holds fixed the dimensionless quantity $rM_{11}$.
We see that the corresponding asymptotic behavior of ${\cal H}_{00}$ is
given by
\begin{equation}
{\cal H}_{00}  \approx \left( \frac{r_0}{r} \right)^7 = \frac{ gN \ell_s^7}
{{\cal H}_4^{1/4}  R^7}.
\end{equation}
As a result, keeping only the order zero term yields an unperturbed value
$R_0$ of $R$ given by
$R_0 \sim (gN)^{-1/4} \ell_s^{-3/4}
{\cal H}_4^{7/16}  R_0^{7/4}$, or
$R_0 \sim (gN)^{1/3} \ell_s
{\cal H}_4^{-1/12}$ in agreement with (\ref{unpert}).

Recall that the effect of the ${\cal H}_{01}$ term is to shift the bubble
by an angle-dependent amount but not to change the size of the bubble.  Due to
the angle-dependence, concepts like the radius $R$ of the bubble also become
angle-dependent.  However, it is the average $\langle R^2\rangle$
of $R^2$ over the bubble
that we expect to compare with expectation values in
the quantum mechanical ground state.    Taking such an average, it is
clear that the term in (\ref{Reqn}) that is linear in ${\cal H}_{01}$ has no
effect on $\langle R^2 \rangle$.  Of course, a shift of the bubble
away from the origin will contribute to $\langle R^2 \rangle$ at second
order, and this effect is governed by the term
$\frac{5}{20}({\cH}_{01})/(4{\cH}_{00})^2$.
This is in agreement with our quantum mechanical calculation where the
effect on $\langle R^2 \rangle$ appeared only at second order in $f$.
The result after averaging,and taking the decoupling limit in the necessary
order of accuracy gives
\be
\label{cgres}
\frac{\langle R\rangle^2 - R_0^2}{R^2_0} \sim f^2{\cal H}_4^{-1/6} (gN)^{2/3},
\ee
in agreement with the quantum mechanical results of (\ref{qmres}).

\begin{theacknowledgments}
The author would like to thank Mark Bowick, Amanda Peet, Joe
Polchinski, and Joel Rozowsky for useful discussions. This work
was supported in part by NSF grant PHY97-22362 to Syracuse
University, the Alfred P. Sloan foundation, and by funds from
Syracuse University.
\end{theacknowledgments}

\end{document}